\documentclass[5p, times]{elsarticle}
\usepackage{geometry}
\usepackage{fleqn}
\usepackage{graphicx}
\usepackage{txfonts}
\usepackage{hyperref}
\usepackage{amssymb}
\usepackage{comment}

\usepackage{lineno,hyperref}
\usepackage{graphics}
\usepackage{algorithm}
\usepackage{blindtext}

\usepackage{scrextend}
\usepackage{graphicx}
\usepackage{mathrsfs}
\usepackage{subcaption}
\usepackage{algpseudocode}
\usepackage{algorithmicx}
\usepackage{color}
\usepackage{amsmath}
\usepackage{epsfig}
\usepackage{multirow}
\usepackage{overpic}
\usepackage{colortbl}
\usepackage{blindtext}
\usepackage{amsmath}
\usepackage{siunitx}
\usepackage{verbatim}
\usepackage[inline]{enumitem}
\usepackage{verbatim}
\usepackage{mathrsfs}

\usepackage{lineno}

\journal{Nuclear instruments and methods in physics research A}

\begin{document}

\begin{frontmatter}

\title{Development of a Data-Driven Method to Simulate the Detector Response of Anti-neutron at BESIII}
\author[First,Second]{Liang Liu\corref{cor1}\fnref{label2}}
	\ead{liangzy@mail.ustc.edu.cn}
\author[First,Second]{Xiaorong Zhou\corref{cor1}\fnref{label2}}
	\ead{zxrong@ustc.edu.cn}
\author[First,Second]{Haiping Peng\corref{cor1}\fnref{label2}}
	\ead{penghp@ustc.edu.cn}

\affiliation[First]{organization={Department of Modern Physics, University of Science and Technology of China},
	postcode={230026},
	city={Hefei},
	country={China}}

\affiliation[Second]{organization={State Key Laboratory of Particle Detection and Electronics},
	postcode={230026},
	city={Hefei},
	country={China}}

\begin{abstract}
	In this paper, a data-driven method to precisely simulate the detector response of the anti-neutron depositing in the Electromagnetic Calorimeter (EMC)
	at BESIII is introduced.
        A large anti-neutron data sample can be selected using the decay
        $J/\psi\to p\bar{n}\pi^{-}$ from the BESIII data sample of 10 billion
        $J/\psi$ events. 
        The detection efficiency for and various observables of
        anti-neutrons interacting in the EMC detector are simulated, taking
        the correlations among the variables into consideration. 
        The systematic uncertainty of this data-driven simulation
        method is determined to be less than 1\% on average.
        This method can be widely applied in physics processes
        that require precise simulation of the detector response of
        anti-neutrons in the EMC.
\end{abstract}




\end{frontmatter}

\section{Introduction}
\label{introduction}

The neutron is one of the primary building blocks of atomic matter in the visible universe.
There is a rich physics program involving neutrons in the final state in electron-positron collider experiments, 
including the study of the electromagnetic form factors of the
neutron via $e^{+}e^{-}\to n\bar{n}$~\cite{Ablikim:2021ogk}, and
measurements of hyperon~($Y$) decays  which 
will eventually produce neutrons. 
Reference~\cite{Yuan:2021yks} discusses a possible new source of neutrons in $J/\psi$ decays,
showing a large potential to study many open problems in particle and nuclear physics.
One further example is the investigation of $CP$ violation in hyperon decays via $J/\psi\to Y\bar{Y}$~\cite{BESIII:2018cnd, BESIII:2021ypr} at BESIII, where the experimental sensitivity is $\mathcal{O}(10^{-2})$ with 1.3~billion $J/\psi$, dominated  by statistical uncertainty.
The BESIII experiment 
has collected a data sample of 10~billion $J/\psi$ events, which can be used to study the decays of hyperons with unprecedented precision~\cite{Li:2016tlt}.  
Using neutral decay modes of hyperons, the statistics will be further improved due to their large branching fractions, {\it e.g.}
$\mathcal{B}(\Lambda \to n\pi^0) = (35.8\pm 0.5)\%$, $\mathcal{B}(\Sigma^+ \to n\pi^+) = (48.31\pm 0.30)\%$~\cite{ParticleDataGroup:2020ssz},
and more independent $CP$ observables can be constructed from the neutral decay parameters. 
Moreover, the $\Sigma^{-}$ hyperon dominantly decays to neutrons, $\mathcal{B}(\Sigma^- \to n\pi^-) = (99.848\pm 0.005)\%$~\cite{ParticleDataGroup:2020ssz}, and its production or decays are hardly discussed in electron-positron collision experiments due to difficulties in neutron reconstruction.

Usually, the anti-particles to neutrons, namely anti-neutrons, are used in the reconstruction instead of neutrons due to their large energy  deposition in the calorimeter 
following an annihilation reaction, which allows for a good discrimination between electromagnetic showers from the annihilation products and photons, measured with the EMC.
However,
a large discrepancy is observed between data and the simulation of the anti-neutron detector response based on {\sc Geant4}~\cite{GEANT4:2002zbu}.
Though many models have been implemented in {\sc Geant4} to describe hadron-nucleon interactions, such as 
quark-gluon string model (QGS), the Fritiof parton model (FTF) or the Bertini and Precompound models, none of them gives a 
satisfactory result for the simulation of anti-neutrons.
A  lot of efforts spent on
correcting the efficiencies of anti-neutron reconstruction in physics analyses~\cite{BESIII:2018cnd,BESIII:2020raf,BESIII:2020uqk},
and the systematic uncertainties caused by the {\sc Geant4}-based simulation are still difficult to determine.

In this paper, we introduce a model-independent data-driven method to simulate the detector response of the anti-neutron at BESIII. 
Both the detection efficiency and various observables of anti-neutrons are sampled from data,
which is independent from the {\sc Geant4} simulation. 
The paper is organized as follows. A brief description of the BESIII experiment is introduced  in Sec.~\ref{sec:besiii}.
The selection of the anti-neutron sample is presented in Sec.~\ref{sec:datasample}.
Section~\ref{sec:algorithm} illustrates the detailed data-driven simulation of the detector response to anti-neutrons in the EMC.
Section~\ref{sec:comparison} gives two comparisons between the simulation and data. Finally, a conclusion and further prospects are presented in Sec.~\ref{sec:summary}.

\section{The BEPCII and the BESIII experiment}
\label{sec:besiii}

The BESIII detector~\cite{BESIII:2009fln} records symmetric $e^+e^-$ collisions 
provided by the BEPCII storage ring~\cite{Yu:2016cof}, which operates with a peak luminosity of \SI{1e33}{cm^{-2}. s^{-1}}
in the center-of-mass energy range from 2.0 to 4.95~GeV.
The cylindrical core of the BESIII detector covers 93\% of the full solid angle and consists of a helium-based
 multilayer drift chamber~(MDC), a plastic scintillator time-of-flight
system~(TOF), and a CsI(Tl) electromagnetic calorimeter~(EMC),
which are all enclosed in a superconducting solenoidal magnet
providing a 1.0~T  magnetic field. The solenoid is supported by an
octagonal flux-return yoke with resistive plate counter muon
identification modules interleaved with steel. 
The charged-particle momentum resolution at $1~{\rm GeV}/c$ is
$0.5\%$, and the specific energy loss ($dE/dx$) resolution is $6\%$ for electrons
from Bhabha scattering.
 The time resolution in the TOF barrel region is 68~ps, while
that in the end cap region is 110~ps. The end cap TOF
system was upgraded in 2015 using multi-gap resistive plate chamber
technology, providing a time resolution of
60~ps~\cite{Li:2012syl,2017The,Cao:2020ibk}.

The EMC at BESIII is designed to precisely measure the deposited energies of neutral particles
 and to provide trigger conditions. 
It is comprised of 6240 CsI(Tl) crystals, arranged in one barrel and two end caps.
The barrel has an inner radius of 94~cm and a length of 275~cm. The end caps have an inner length
of 50~cm and are placed 138~cm from the collision point.
The angular coverage is within the polar angle range of $|\cos\theta|<0.82$ for the barrel and $0.83 < |\cos\theta| < 0.93$ for the end caps. 
The EMC measures the energies of electrons and photons with
an energy resolution of $2.5\%$ ($5\%$) at $1$~GeV in the barrel (end cap)
region; the position is measured with a resolution of $\SI{0.6}{\centi\metre}/\sqrt{E~(\rm GeV)}$.

Using this detector setup, anti-neutrons can be detected by the EMC with a relatively high efficiency. 
The anti-neutrons are likely to annihilate in the EMC~\cite{ASTRUA2002209} and produce several secondary particles. 
The largest energy deposition ($E_{\bar{n}}$) can be as high as 2~GeV.  
Other variables used to characterize  showers from the anti-neutron in the EMC include the number of hits ($H_{\bar{n}}$) from the 
primary shower and 
the second moment ($S_{\bar{n}}$) of the shower, defined as $S_{\bar{n}} = \sum_i E_i r_i^2 / \sum_i E_i$, with $E_i$ the energy deposited in the 
$i^\text{th}$ crystal of the shower and $r_i$ the distance from the center of that crystals to the center of the shower.

\section{Anti-neutron samples}
\label{sec:datasample}

\subsection{Selection of $J/\psi \to p \bar{n} \pi^-$}
\label{subsec:jpsipnbarpi}

To study the performance of anti-neutrons in the EMC, a large sample of anti-neutrons with high purity is needed.
The process $J/\psi \to p \bar{n} \pi^-$ has several advantages for the study of anti-neutron properties from data:
a large amount of anti-neutron events can be selected from 10 billion $J/\psi$ events thanks to a
fraction of $\mathcal{B}(J/\psi \to p \bar{n} \pi^-)=(2.12\pm0.09)\times10^{-3}$~\cite{ParticleDataGroup:2020ssz};
the momentum of the anti-neutrons covers a large range from above 0 to 1.2~GeV/$c$ in the whole $4\pi$ angular acceptance;
the anti-neutron signal can be determined precisely with low backgrounds by
reconstructing the four-momenta of $p$ and $\pi^{-}$ and studying the recoil side of $p\pi^{-}$.

\begin{figure}[htbp!]
 \centering
      \includegraphics[width=0.45\textwidth]{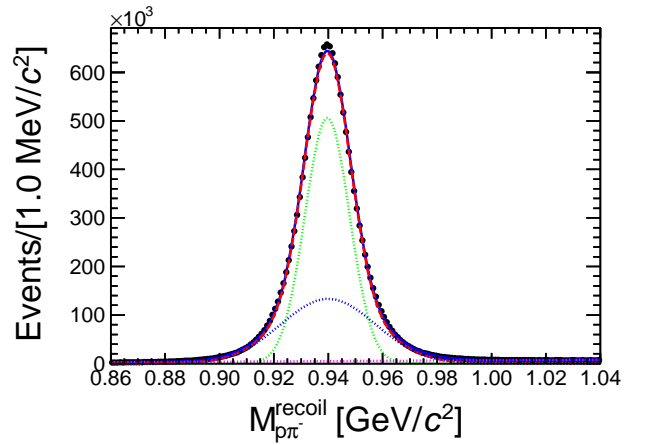}
	\caption{ Distribution the invariant mass recoiling against the $p\pi^-$ system.  The black dots with error bars are data;
	the blue line indicates the total fit; the long dashed red line represents the signal contribution described by a
         double-Gaussian function, which is comprised of the green and blue short dashed lines; the dashed magenta line 
	 indicates the background described by a linear function.}
\label{fig:ppi_recoil}
\end{figure}

The charged tracks are reconstructed from hits in the MDC.
The polar angles of charged tracks are required to fulfill $|\cos \theta| < 0.93$.
The distance of closest approach to the interaction point of each charged track
must be within $\pm5$~cm along the beam direction ($V_z$)
and be within $0.5$~cm in the plane perpendicular to the beam axis ($V_{xy}$). 
This vertex requirement is applied to suppress beam-associated backgrounds. 
Two tracks with opposite charge are required in each event.
Any events with additional charged tracks within $|V_z| < 30$~cm and $V_{xy}< 10$~cm are
rejected to suppress the background mainly caused by anti-neutron annihilation within the beam pipe material.
The combined information from $dE/dx$  and TOF is used to calculate particle identification (PID) probabilities 
for the pion, kaon and proton hypotheses, respectively, and the particle type with the highest probability
is assigned to the corresponding track.
In each event, exactly one $p^+$ and one $\pi^{-}$ are required. 
A vertex fit is performed for the two selected tracks to improve the momentum resolution and the corresponding
$\chi^{2}$ of vertex fit is required to be less than 30. 
After the above selection, the invariant mass of the system recoiling against $p\pi^{-}$,  $M_{p\pi^{-}}^{\rm recoil}$, is shown
in Fig.~\ref{fig:ppi_recoil},
where a clear peak is observed in data that indicates the signal from the anti-neutron.

A large inclusive sample of about 10 billion simulated events is used to estimate possible background contributions,
where the $J/\psi$ resonance is generated by {\sc kkmc}~\cite{Jadach:1999vf,Jadach:2000ir} subsequent decays with known branching 
fractions are generated by {\sc EvtGen}~\cite{Lange:2001uf}, and the remaining unmeasured decays are 
generated by {\sc LundCharm}~\cite{Chen:2000tv}.  
Studies of the inclusive MC simulation indicate that very few background events can survive after application of 
above selection  criteria, and the main background contribution comes from 
the process $J/\psi \to p\bar{n}\pi^-\gamma$, which does not peak in the signal region of $M_{p\pi^{-}}^{\rm recoil}$.
To obtain the signal yield of $J/\psi\to p\bar{n}\pi^{-}$, 
a binned maximum likelihood fit is performed on the $M_{p\pi^{-}}^{\rm recoil}$ distribution,
where the signal is described by a double-Gaussian shape function, and 
the background is described by a first-order  Chebyshev polynomial. 
The fit result is shown in Fig.~\ref{fig:ppi_recoil}
and yields $(1.6213\pm0.0004)\times10^{7}$
signal events. The error is statistical only. 
The background contribution in the signal region $(0.903,0.977)~{\rm GeV}/c^2$ is about 2.1\% and flat. 
The results of the fit are used to determine the amount of background events as a pre-study for the following sideband analysis within $(1.01,1.03)~{\rm GeV}/c^2$.

\subsection{Anti-neutron selection}
Neutral shower candidates are reconstructed from isolated clusters in the EMC crystals. Efficiency and
energy resolution are improved by including the energy deposited in the nearby TOF counters. 
A minimum energy of 25~MeV in barrel region ($|\cos\theta| < 0.8$) or
50~MeV in end caps region ($0.86 < |\cos\theta| < 0.92$) is required for a good neutral shower.
The most energetic among the good neutral showers is assigned to
be the anti-neutron candidate. With this requirement, the anti-neutron shower can be selected with a purity over 99\% from data,
which is estimated by matching the position of the selected shower with that predicted by the $p\pi^{-}$ recoil system. 

\begin{figure*}[htbp!]
 \centering
  \mbox
  {
  \begin{overpic}[width=0.3\textwidth]{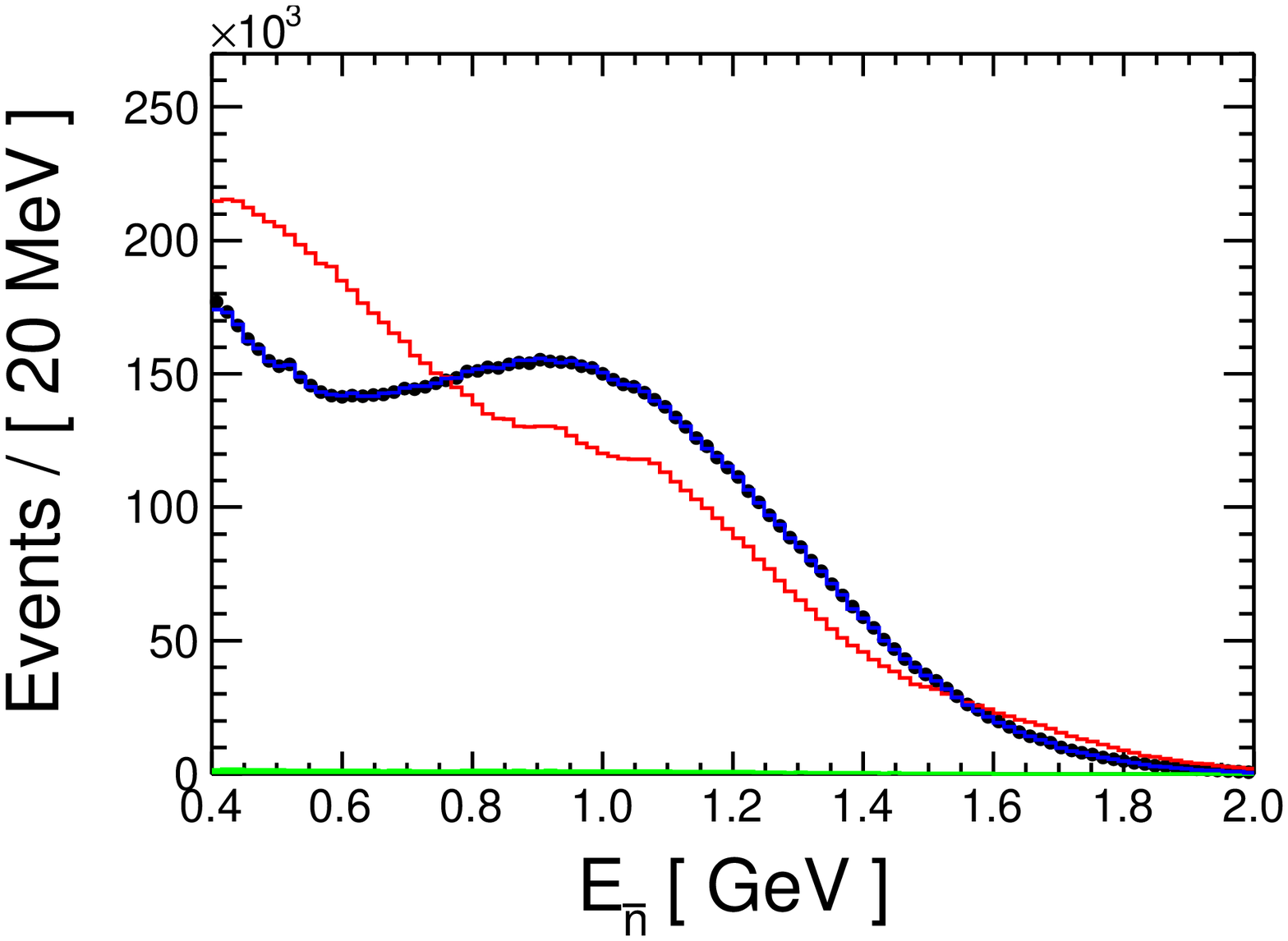}
  \put(80,55){\small{(a)}}
  \end{overpic}
  }
\mbox
{
  \begin{overpic}[width=0.3\textwidth]{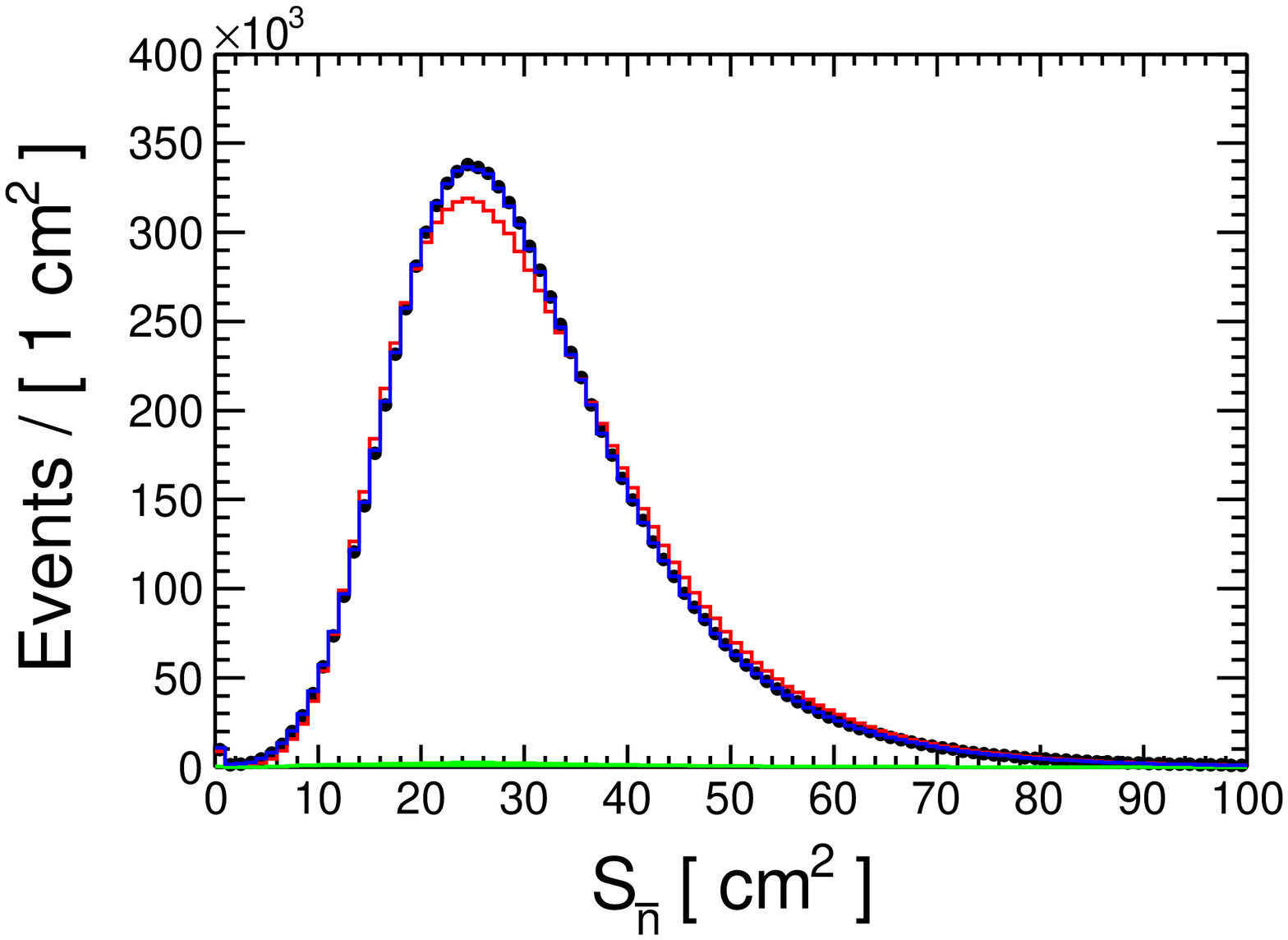}
  \put(80,55){\small{(b)}}
  \end{overpic}
}
\mbox
{
  \begin{overpic}[width=0.3\textwidth]{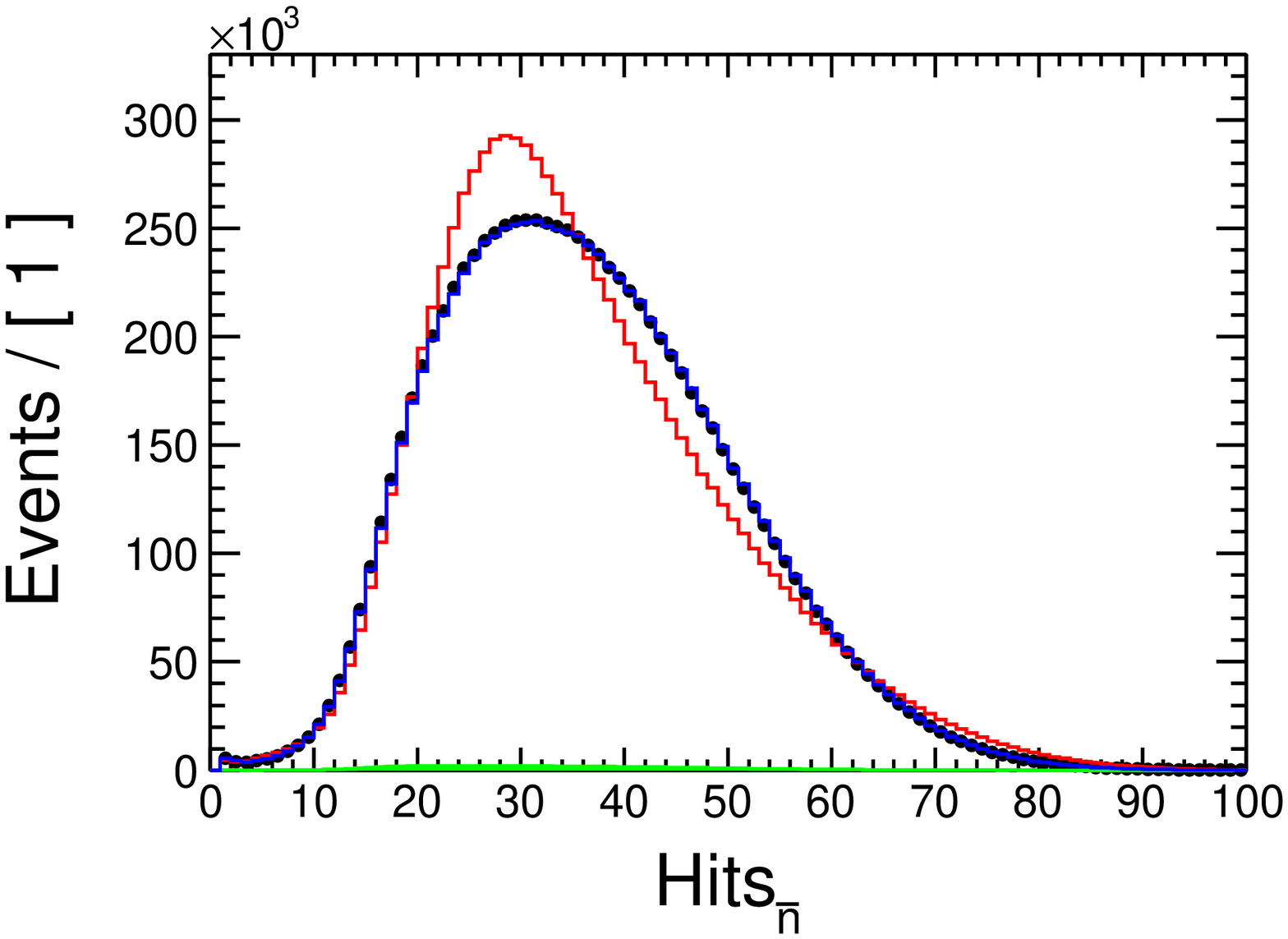}
  \put(80,55){\small{(c)}}
  \end{overpic}
  }
\caption{Distributions of deposited energy (a), second moment (b), number of hits (c) from $\bar{n}$ candidates.
	The black dots with error bars show data; the red lines indicate weighted PHSP MC which is {\sc Geant4}-based;
	The blue lines are {\sc Geant4}-independent simulation of anti-neutron properties introduced in this paper;
	the green areas represent the sideband region.
}
\label{fig:showerenergy}
\end{figure*}

Several distributions of anti-neutron showers  are 
investigated with the selected anti-neutron sample as shown in Fig.~\ref{fig:showerenergy}.
The definitions of signal and sidebands are the same as in Sec.~\ref{subsec:jpsipnbarpi}.
A further selection criterion on the deposited energy $E_{\bar{n}}$ from the primary anti-neutron shower, 
$E_{\bar{n}}>0.4$~GeV,
is applied to distinguish the signal showers from background.
The {\sc Geant4}-based signal MC for $J/\psi\to p\bar{n}\pi^{-}$ is simulated with a uniform phase space~(PHSP) generator and
is used for a comparison with data.
To have a fair comparison between data and  signal MC, a weight depending on the momentum and polar 
angle of the anti-neutron is applied to the signal MC.
A large discrepancy between simulated and measured  anti-neutrons is observed as shown in Fig.~\ref{fig:showerenergy}, which indicates
that the simulation of anti-neutrons based on {\sc Geant4} is  unreliable.
In many analyses, various requirements 
 on $E_{\bar{n}}$, $S_{\bar{n}}$ and $H_{\bar{n}}$ are applied on the anti-neutron
to further  suppress background events. 
The large discrepancy in these distributions will cause a large error for the detection
efficiency. Therefore, the simulation performance of anti-neutrons in the EMC needs to be improved. In the following, we introduce
a {\sc Geant4}-independent way  to simulate an anti-neutron detector response via a data-driven method.

\section{Simulation of the detector response to anti-neutrons}

\label{sec:algorithm}

The detection efficiency and  observables in the EMC for anti-neutrons can be treated separately 
in the {\sc Geant4}-independent simulation.
Firstly, a detection efficiency for anti-neutrons is obtained from the selected
$J/\psi\to p\bar{n}\pi^{-}$ sample, as determined by
the number of signal events in  $M_{p\pi^{-}}^{\rm recoil}$ before and after the shower selection as described in 
Sec.~\ref{sec:datasample} including any further requirements on the anti-neutron shower. The efficiency  of
anti-neutron from MC simulation can then be sampled with the
Accept-Reject Sampling method~\cite{2004Generalized}.
Secondly, various observables are simulated according to the distributions from data via an inverse transform sampling method~\cite{Caflisch1998Monte},
including the characteristic variables which describe the shower of the anti-neutron, {\it e.g.}  $E_{\bar{n}}$, $S_{\bar{n}}$ and $H_{\bar{n}}$,
and kinematic variables, {\it e.g.} polar angle $\theta_{\bar{n}}$ and azimuth angle $\phi_{\bar{n}}$, as well as their errors.

\subsection{Detection efficiency}

The detection efficiency of the anti-neutron for any selection criteria can be well expressed in terms of 
momentum $p_{\bar{n}}$ and polar angle $\cos\theta_{\bar{n}}$.
The efficiency surface is determined by 

\begin{equation} \label{eq:efficiency_surface}
\varepsilon(p_{\bar{n}}, \cos\theta_{\bar{n}}) = \frac{f'(p_{\bar{n}}, \cos\theta_{\bar{n}})} { f (p_{\bar{n}}, \cos\theta_{\bar{n}})},
	\end{equation}
where $f$ and $f'$ stand for the number of signal events in a given $p_{\bar{n}}$ and $\cos\theta_{\bar{n}}$  range
before and after applying the selection criteria, respectively.
For the 2D efficiency matrix, the momentum $p_{\bar{n}}$ is divided into 50 bins in the range (0, 1.2)~GeV/$c$,
and the range of the polar angle $\cos\theta_{\bar{n}}$ is divided into 36 bins for $|\cos\theta_{\bar{n}}|<0.72$ and 84 bins for $0.72<|\cos\theta_{\bar{n}}|<1$.
The number of signal events is determined by counting the events  of $M_{p\pi^-}^{\rm recoil}$ in the 
corresponding range of $p_{\bar{n}}$ and $\cos\theta_{\bar{n}}$ in data
after subtracting the background events as described in Sec.~\ref{sec:datasample}.

\begin{figure}
\centering
	\includegraphics[width=0.45\textwidth]{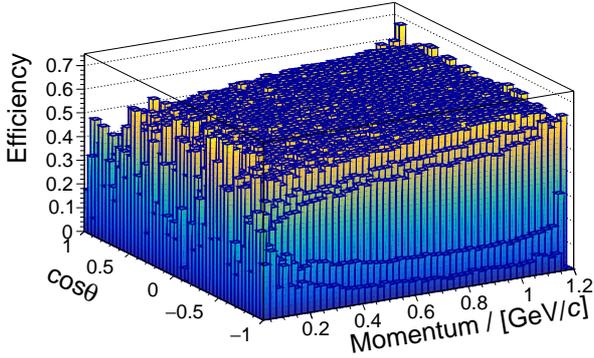}
	\caption{ Example of an anti-neutron efficiency surface as a function of momentum and $\cos\theta_{\bar{n}}$, with the further selection criterion $E_{\bar{n}}>0.4$~GeV.
	}
\label{fig:pcoseff}
\end{figure}

An example distribution of the anti-neutron efficiency as a function of  $p_{\bar{n}}$ and $\cos\theta_{\bar{n}}$ obtained from data is shown in Fig.~\ref{fig:pcoseff}.
Such a model independent, data-driven efficiency $\varepsilon(p_{\bar{n}},\cos\theta_{\bar{n}})$
can be used 
to replace the efficiency in {\sc Geant4}-based MC by adopting the Accept-Reject Sampling method~\cite{2004Generalized}.
The framework for simulating the detection efficiency is presented in Algorithm~\ref{al:eff}.

	\begin{algorithm}
	\caption{Simulation of detection efficiency of the anti-neutron. Event accepted means that the corresponding event is detected for a given set of selection criteria, otherwise that event is lost. }
	\label{al:eff}
	\begin{algorithmic}
	\ForAll{events (after preliminary selection criteria)} 
	\State Extract the momentum  $p_{\bar{n}}$ and polar angle $\cos\theta_{\bar{n}}$ of anti-neutron from generator;
	\State Determine detection efficiency of anti-neutron  $\varepsilon_{i}$ according to Eq.~\ref{eq:efficiency_surface};
	\State Generate $u_{i} \sim \mathcal{U}(0, 1)$ (the uniform distribution over the unit interval).
	\If {$u_{i} \leq \varepsilon_{i}$}
	\State Anti-neutron has information in EMC, event accepted.
	\Else
	\State No information of anti-neutron is left in EMC, event rejected.
	\EndIf
	\EndFor
	\end{algorithmic}
	\end{algorithm}

In Algorithm~\ref{al:eff}, the detection efficiency for an anti-neutron is sampled according to an efficiency $\varepsilon_i$ in 
a  $(p_{\bar{n}}, \cos\theta_{\bar{n}})$ bin of the anti-neutron as expressed above.
The intrinsic uncertainty of $\Delta\varepsilon_{i}$
 is given by summing  statistical and systematic uncertainties in quadrature, which will be used to derivate the systematic
 uncertainties of this algorithm.
 The systematic uncertainties of the detection efficiency mainly come from the estimation of the background contribution,
where alternative fits are performed on  $M_{p\pi^-}^{\rm recoil}$  with a second-order polynomial for background,
or the usage of a different fitting range.
The resulting differences on the background estimation are taken as the systematic uncertainty.
The average systematic uncertainties from the background shape and fit range are 0.3\% and 0.03\%, respectively.

\subsection{Anti-neutron observables in EMC}

Typically, there are two kinds of observables for anti-neutrons to be considered in physics analyses.
The first group comprises the characteristic variables  used to distinguish anti-neutron showers from photon showers, such as
 $E_{\bar{n}}$, $S_{\bar{n}}$ and $H_{\bar{n}}$ {\it etc.} These applications can be found in Ref.~\cite{Ablikim:2021ogk,BESIII:2018cnd}.
The second group contains the kinematic-related variables which can be involved in kinematic fitting~\cite{Yan:2010zze}, such as in Ref.~\cite{BESIII:2020raf,BESIII:2020uqk}.
For kinematic fits including anti-neutrons, the energy information is missing since anti-neutrons do not actually
deposit all their energy in the EMC. Instead, the position information, $\theta_{\bar{n}}$ and $\phi_{\bar{n}}$,
and the corresponding errors, $\delta \theta_{\bar{n}}$ and $\delta \phi_{\bar{n}}$, are implemented in 
the kinematic fit and therefore need to be simulated precisely. The framework for the simulation of such
kind of observables is presented in Algorithm~\ref{al:cdf}.

\begin{algorithm}
  \caption{Simulation of observables for anti-neutron in EMC, taking $E_{\bar{n}}$ as an example. }
  \label{al:cdf}
  \begin{algorithmic}
    \ForAll{anti-neutrons with information in EMC from Algorithm~\ref{al:eff}} 
    \State Generate $u \sim \mathcal{U}(0, 1)$ (the uniform distribution over the unit interval).
    \State Find the inverse of the desired CDF, {\it e.g.} $E^{-1}(x)$.
    \State Calculate $E_{\bar{n}}$ by  $E_{\bar{n}} = E^{-1}(u)$.
    \EndFor
  \end{algorithmic}
\end{algorithm}

In Algorithm~\ref{al:cdf}, an inverse transform  sampling method is performed to simulate observables,
where the cumulative distribution functions~(CDFs) are calculated from the normalized distributions of 
the anti-neutrons observables in data. 
The CDF is a monotonly rising function as shown in Fig.~\ref{fig:Ecdf} for $E_{\bar{n}}$ as an example.
With a given random number from 0 to 1, one can determine the corresponding $E_{\bar{n}}$ value. 
This method is widely used for the sampling of a distribution which cannot be easily parameterized.  

Unlike the detection efficiency of the anti-neutron which is only related to 
the momentum  $p_{\bar{n}}$ and polar angle $\cos\theta_{\bar{n}}$, there are correlations among the 
observables. In the following, we describe the details about the simulation of 
characteristic variables  and kinematic-related variables of the anti-neutron.

\begin{itemize}
\item {\bf Characteristic variables of the anti-neutron}

The distributions of  characteristic variables, $E_{\bar{n}}$, $S_{\bar{n}}$ and $H_{\bar{n}}$, are dependent on the  momentum $p_{\bar{n}}$ and the polar angle $\cos\theta_{\bar{n}}$ of the anti-neutron. 
		The momentum range is divided into 50 bins for (0, 1.2)~GeV/$c$, and the polar angle is divided into 50 bins within $|\cos\theta_{\bar{n}}|<1$.
The distributions of simulated observables by means of Algorithm~\ref{al:cdf} are consistent with these from data, as shown in Fig.~\ref{fig:showerenergy}.\\

\begin{figure}[htbp!]
	\centering
	\mbox
	{
		\begin{overpic}[width=0.45\textwidth]{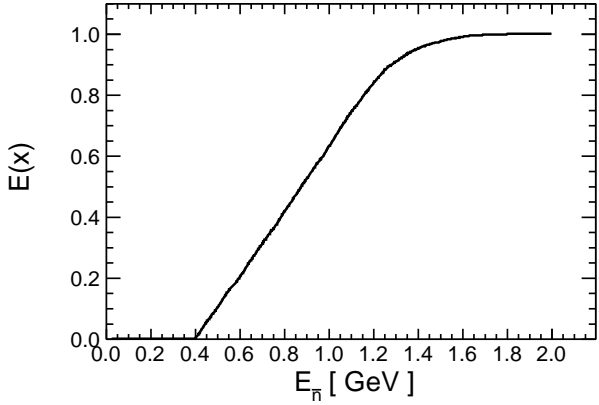}
			\put(80,55){}
		\end{overpic}
		}
		\caption{ Cumulative distribution function (CDF) of $E_{\bar{n}}$ as an example in the 30th bin of $p_{\bar{n}}$ and 25th bin of $\cos\theta_{\bar{n}}$.
	A random variable with distribution $E(x)$ can be calculated from $E^{-1}(u)$ by a random number $u$ generated from a uniform distribution
	in the interval [0, 1].
		}
		\label{fig:Ecdf}
\end{figure}

\item {\bf Kinematic variables of the anti-neutron}

	The kinematic fit including an anti-neutron uses the spatial information $\theta_{\bar{n}}$, $\phi_{\bar{n}}$ 
	measured with the EMC and their error matrix while the deposited energy is a free parameter.  
	The reconstructed $\theta_{\bar{n}}$ and $\phi_{\bar{n}}$ are determined by the spatial resolution
	of the anti-neutron at EMC, \emph{i.e.} the opening  angle  $\Delta\alpha_{\bar{n}}$
	between the position vectors of the anti-neutron shower in the EMC $\vec{V}_{\bar{n}}$ and 
	its direction as simulated on the generator level $\vec{V}_{\rm gen}$.
	A good agreement on $\Delta\alpha_{\bar{n}}$ between data and MC are necessary.
	Figure~\ref{fig:corr_energy} shows the comparison between data and Geant4-based simulation for $\Delta\alpha_{\bar{n}}$, 
	where a large discrepancy is observed. 

	\begin{figure}[htbp]
		\centering
		\mbox
		{
			\begin{overpic}[width=0.45\textwidth]{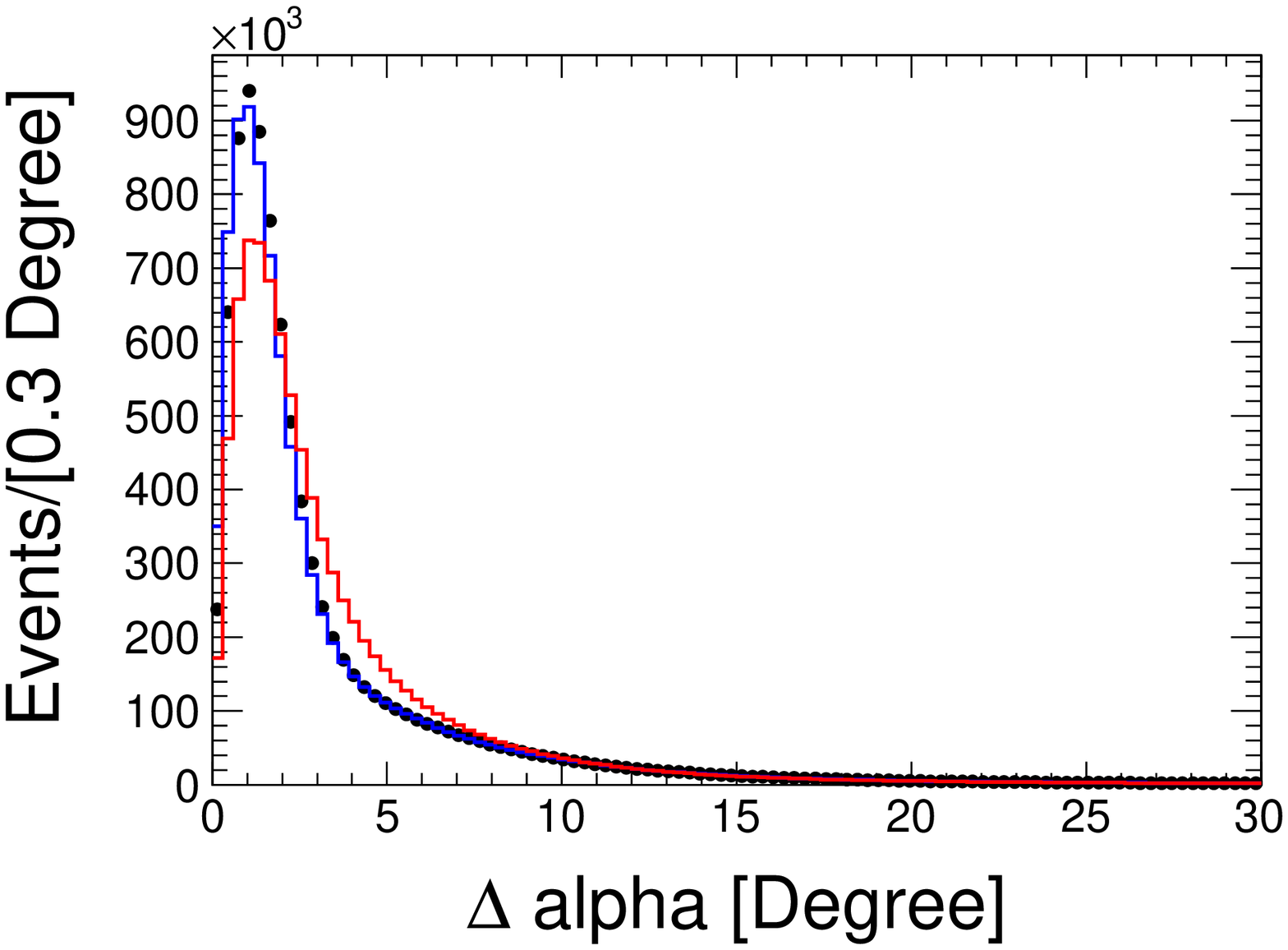}
				\put(80,55){\small{(a)}}
			\end{overpic}
		}
		\mbox
		{
			\begin{overpic}[width=0.45\textwidth]{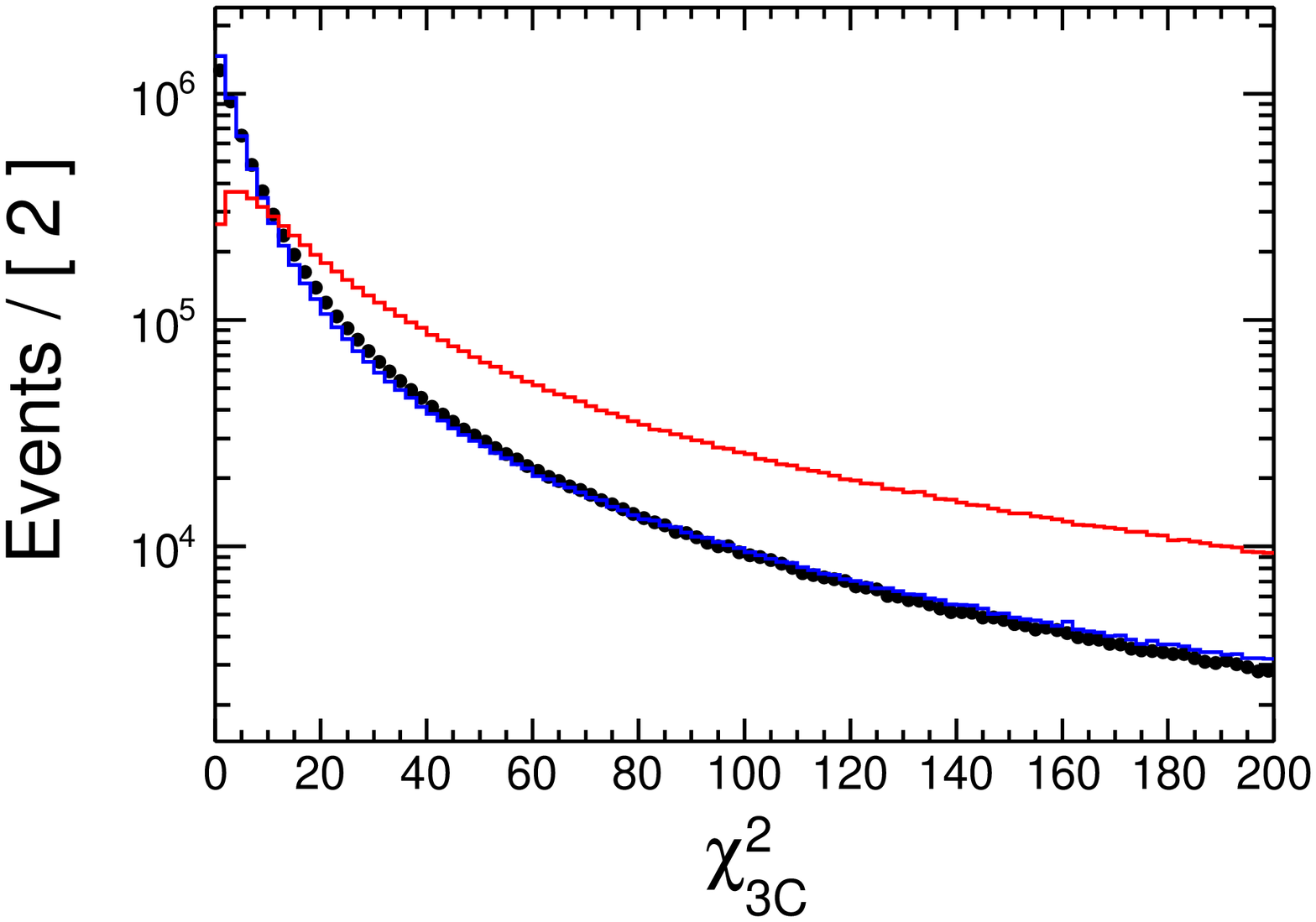}
				\put(80,55){\small{(b)}}
			\end{overpic}
		}
		\caption{ 
		(a) Distribution of the opening angle $\Delta\alpha_{\bar{n}}$
		between the position of the anti-neutron shower in the EMC and its direction on generator level;
		(b)  $\chi^2$ distributions from the 3C kinematic fit.
		The black dots with error bar are results from data. The red lines
		are {\sc Geant4}-based simulation. The blue lines are {\sc Geant4}-independent simulation. }
		\label{fig:corr_energy}
	\end{figure}

	The distribution of $\Delta\alpha_{\bar{n}}$ is relevant for  the momentum $p_{\bar{n}}$, polar angle $\cos\theta_{\bar{n}}$ and the deposited energy $E_{\bar{n}}$. 
	As a consequence, a 3D sampling is necessary for the simulation of $\Delta\alpha_{\bar{n}}$.
	To guarantee that the number of entries in most bins are larger than $2\times10^{3}$,
	the momentum range  $p_{\bar{n}}$ is divided into 12 bins for (0, 1.2)~GeV/$c$; the polar angle range  $\cos\theta_{\bar{n}}$ 
	is divided into 20 bins
	from -1 to 1; the deposited energy $E_{\bar{n}}$ is divided into 10 bins for (0, 1.0)~GeV, 2 bins from 1.0 to 1.4~GeV and one bin for $(1.4, 2.0)$~GeV. With
	the application of Algorithm~\ref{al:cdf} for the simulation of $\Delta\alpha_{\bar{n}}$,
	a good agreement is achieved as shown in Fig.~\ref{fig:corr_energy}.

	The polar angle $\theta_{\bar{n}}$ and azimuthal angle $\phi_{\bar{n}}$  of the anti-neutron can then be calculated from  $\Delta\alpha_{\bar{n}}$, by 

	\begin{subequations} \label{eq:samplethetaphi}
		\label{position}
		\begin{align}
			\theta_{\bar{n}} & =  \sin\beta \cdot \Delta\alpha_{\bar{n}}+ \theta_{\rm gen}  \label{eq:sampledtheta},  \\
			\phi_{\bar{n}}    &=  \frac{\cos \beta}{\sin\theta_{\rm gen}}\cdot \Delta\alpha_{\bar{n}}  + \phi_{\rm gen} \label{eq:sampledphi},
		\end{align}
	\end{subequations}
	where $\theta_{\rm gen}$ and $\phi_{\rm gen}$ are the polar and azimuthal angles of the anti-neutron on the generator level,
	and $\beta$ is a uniform random number in the range between 0 to $2\pi$. \\

\item {\bf  Error matrix of the anti-neutron}

	For the kinematic fit, an initial $2\times2$ error matrix  $V(\theta_{\bar{n}},\phi_{\bar{n}})$ of 
	the anti-neutron is needed as input for further  iterations 
	and for calculating the $\chi^{2}$ value.  As there is no correlation considered between 
	$\theta$ and $\phi$, only the diagonal elements of the error matrix are non-zero.
	By default, the error matrix of the EMC shower is obtained by assuming an electromagnetic 
	shower from a photon or electron. However, the error matrix of the anti-neutron is very different 
	since larger error matrix elements are expected due to its wider shower shape.
	Therefore, the error matrix  elements should be re-simulated for the kinematic fit for both data and MC.
	Two diagonal matrix elements,  $\Delta\theta_{\bar{n}} = \theta_{\bar{n}} -\theta_{\rm gen}$ and $\Delta\phi_{\bar{n}} = \phi_{\bar{n}}-\phi_{\rm gen}$,
	are obtained from Eq.~\ref{position}.
	The resolutions depend on $E_{\bar{n}}$ and $\theta_{\bar{n}}$, and the corresponding 2D sampling is performed with
	$E_{\bar{n}}$ divided into 20 bins from 0 to 2.0 GeV and  $\cos\theta_{\bar{n}}$  divided into 50 bins within $|\cos\theta_{\bar{n}}|<1.0$. 

	As an example, the kinematic fit is performed for $J/\psi\to p\bar{n}\pi^{-}$, including the spatial information 
	of the anti-neutron. It is denoted as a 3C kinematic fit, with four-momentum constraint and one 
	free parameter, the energy deposition $E_{\bar{n}}$ of the anti-neutron. The $\chi^{2}$ distributions
	from the {\sc Geant4}-based simulation and the {\sc Geant4}-independent simulation are shown in Fig.~\ref{fig:corr_energy}(b).

\end{itemize} 

\subsection{Effects of the detector resolution}

There is a small gap between the barrel and end caps in EMC. The edge effect due to spatial resolution, which can lead to migration of anti-neutron showers 
from the barrel to the end caps, or vice versa, needs to be considered.

\begin{figure*}[hpt]
 \centering
  \mbox
  {
  \begin{overpic}[width=0.33\textwidth]{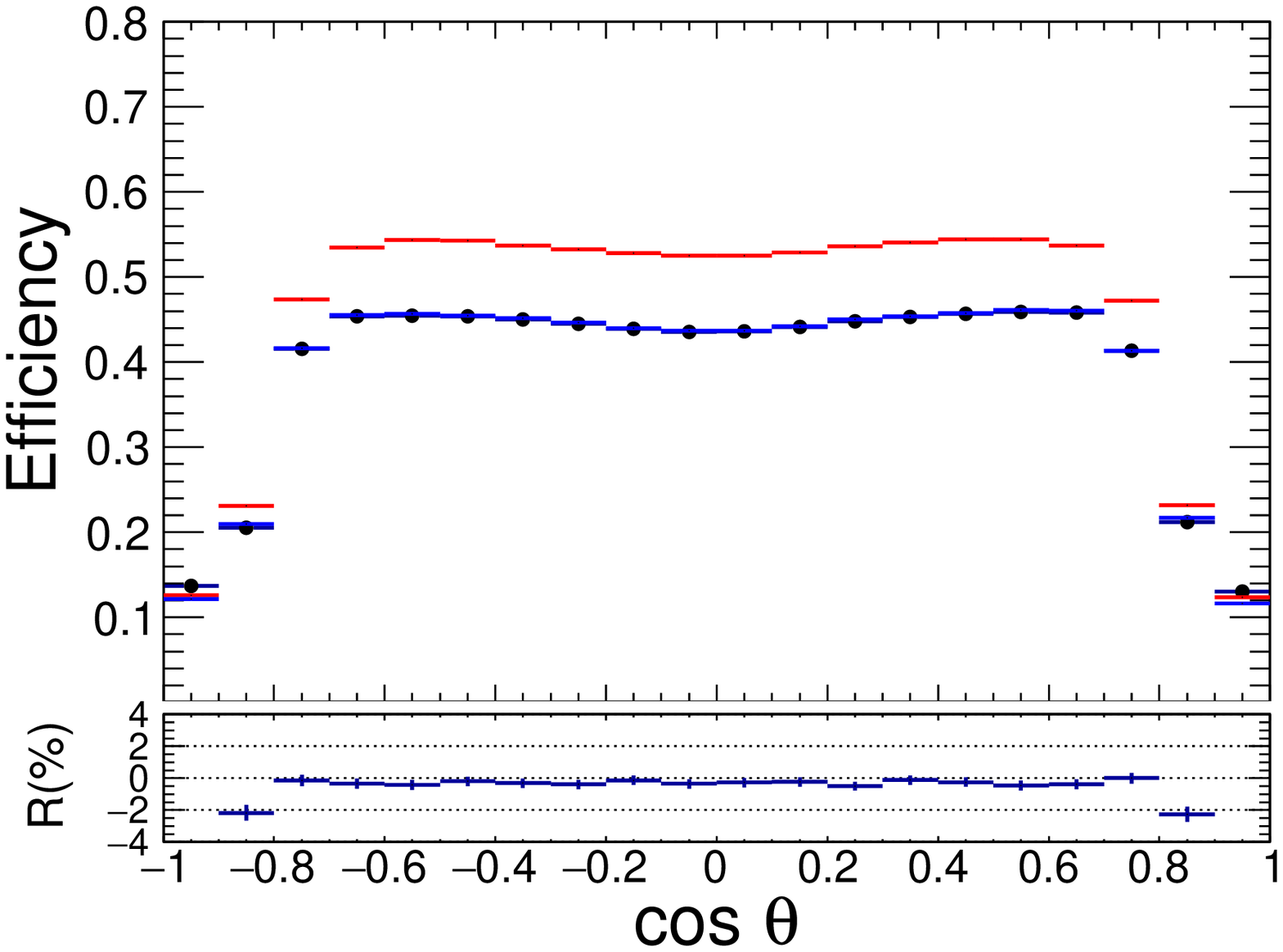}
  \put(80,60){\small{(a)}}
  \end{overpic}
  }
  \mbox
  {
  \begin{overpic}[width=0.33\textwidth]{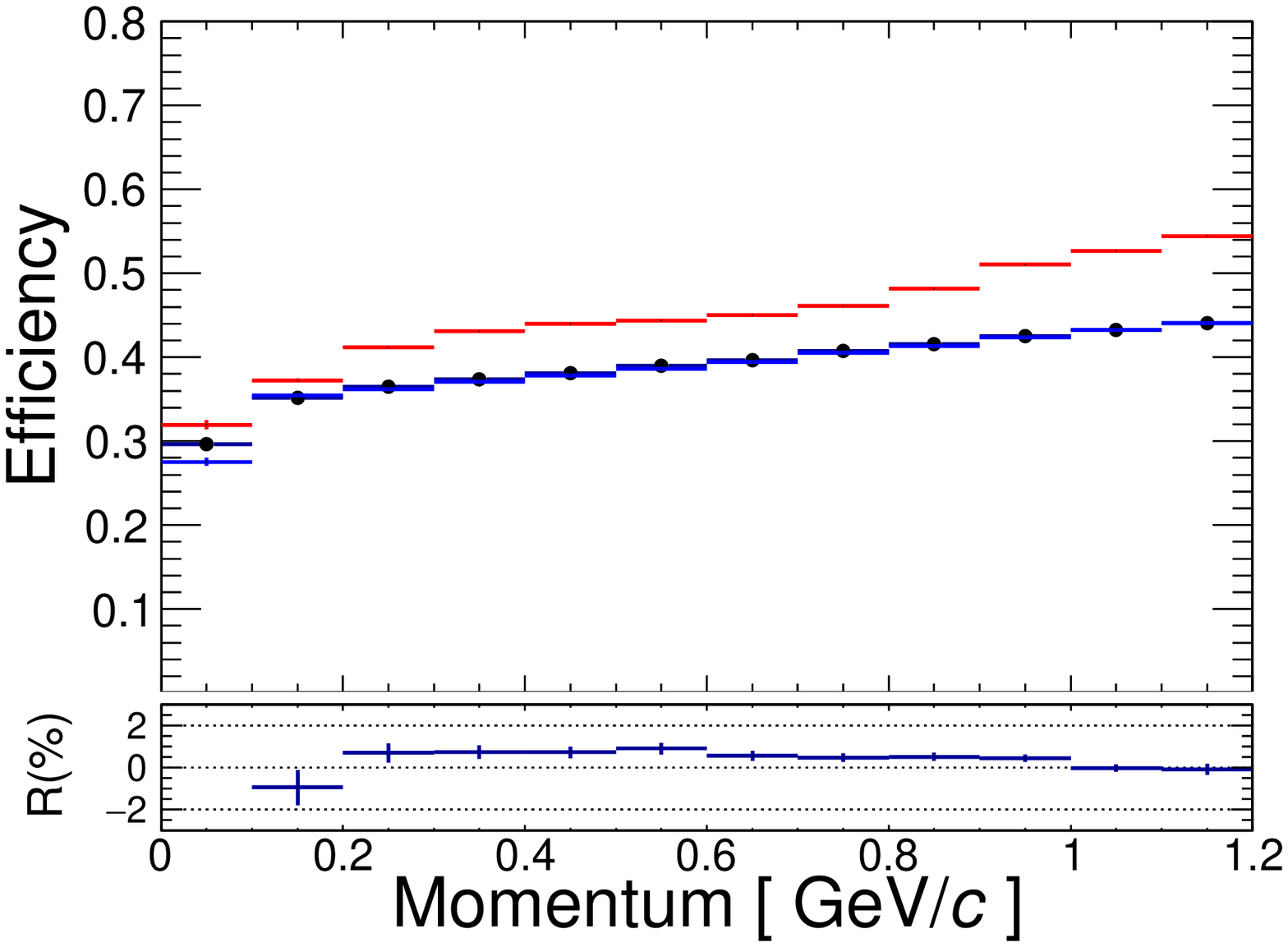}
  \put(80,60){\small{(b)}}
  \end{overpic}
  }\\
  \mbox
  {
  \begin{overpic}[width=0.33\textwidth]{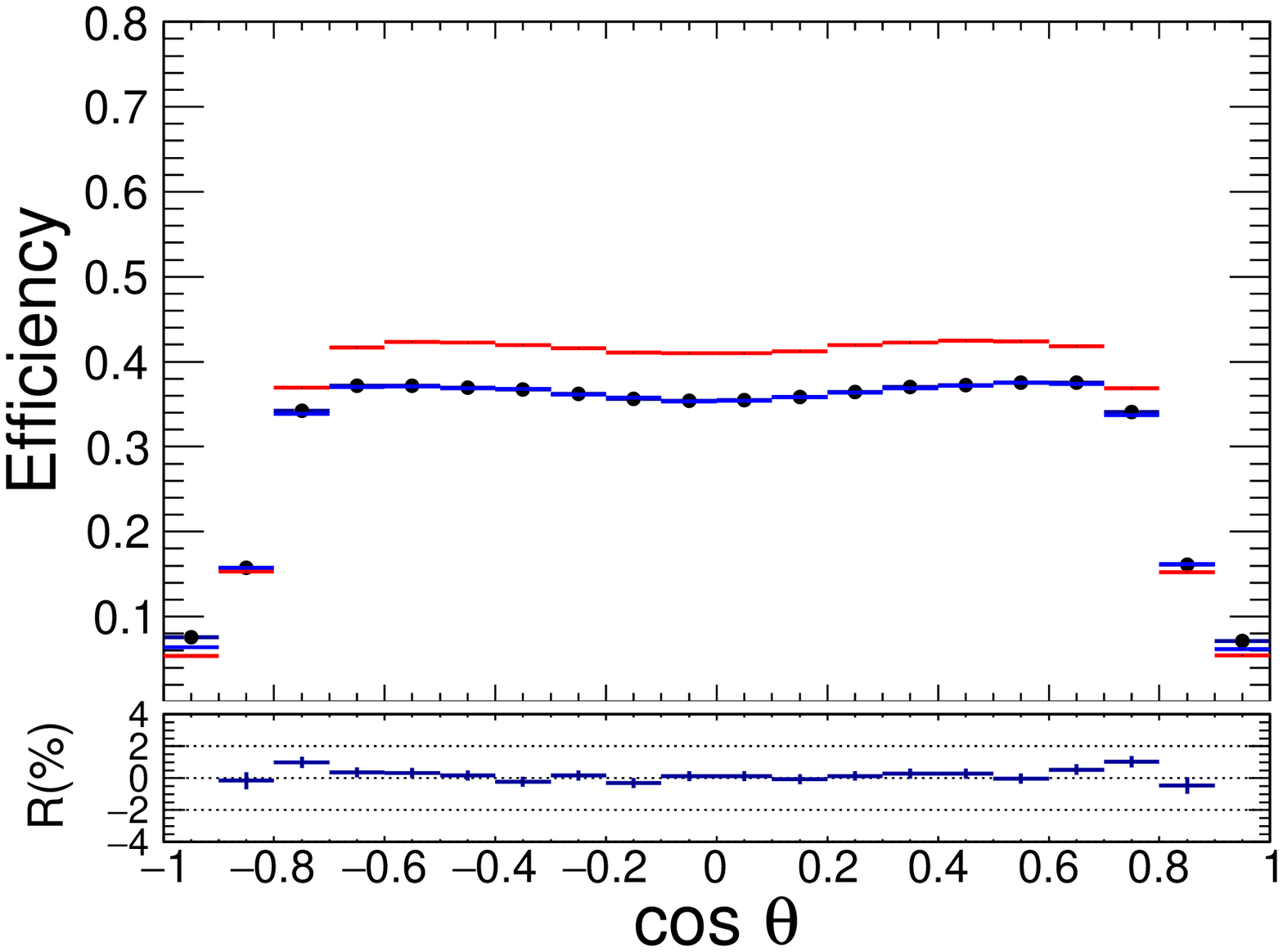}
  \put(80,60){\small{(c)}}
  \end{overpic}
  }
  \mbox
  {
  \begin{overpic}[width=0.33\textwidth]{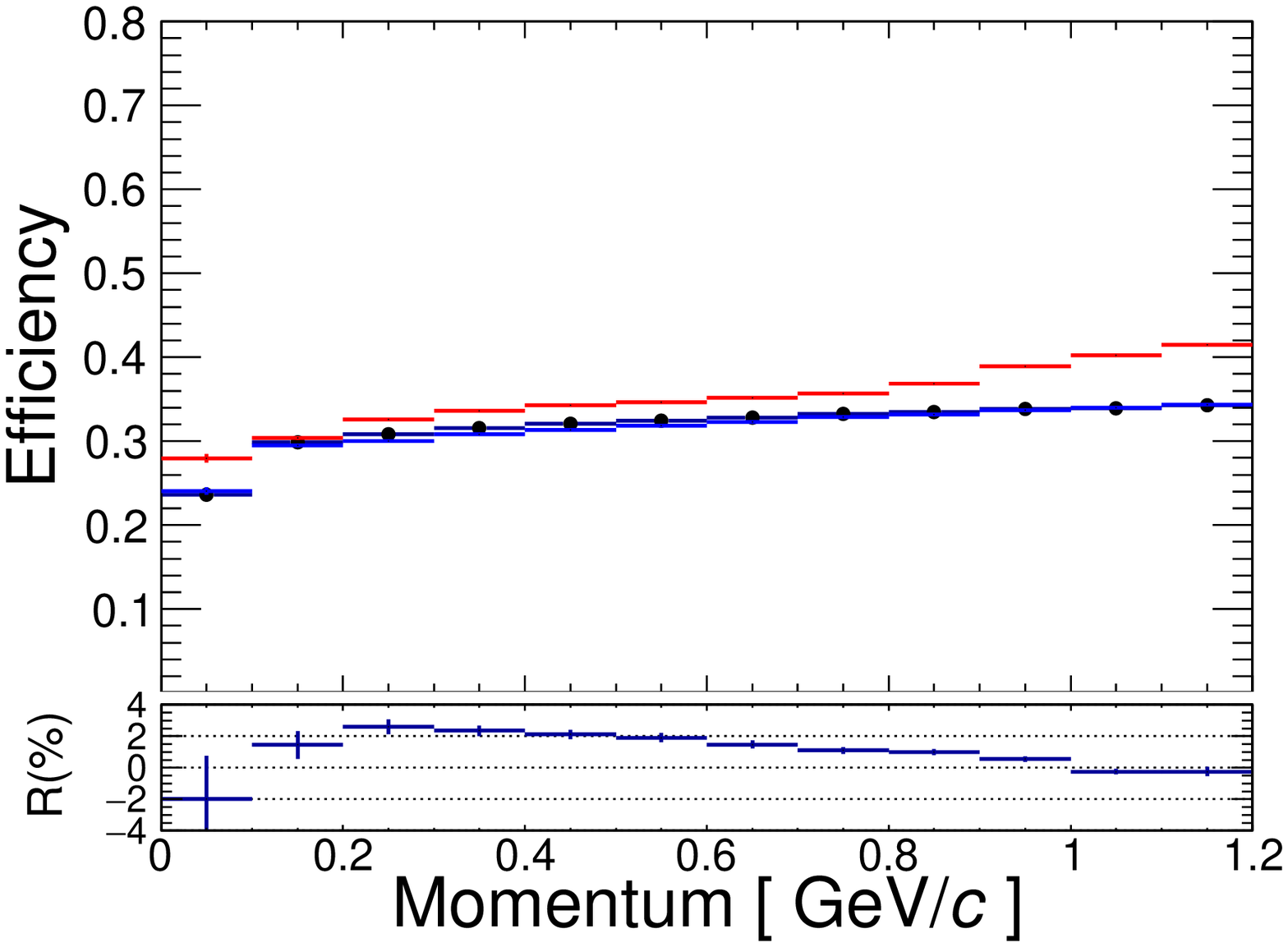}  
\put(80,60){\small{(d)}}
  \end{overpic}
  }
\caption{Comparison of the efficiency of $\bar{n}$ selection between data and simulated MC with respect to $\cos\theta_{\bar{n}}$ and  $p_{\bar{n}}$. The divergence
	between the data efficiency and data-driven simulation one is defined as $R=\varepsilon_{\rm  data}/\varepsilon_{\rm MC}  - 1$.
(a) and (b) are efficiency distributions after the selection of the anti-neutron showers under the conditions that  $E_{\bar{n}}>0.48$~GeV, $S_{\bar{n}}>18$~cm$^{2}$ and $H_{\bar{n}}>20$.
(c) and (d) are efficiency distributions with the additional kinematic fit.
 Black dots with error bars show data;  red lines indicate weighted PHSP simulation which is {\sc Geant4}-beased;  blue lines represents the {\sc Geant4}-independent simulation.
}
\label{fig:efficiency}
\end{figure*}

A simplified  unfolding method is used to correct the edge effect along $\cos\theta$ near the gaps. 
In each momentum bin, a conditional probability $R_i$ is defined as 

\begin{equation}
	\begin{aligned}
		R_{i} &= P(\text { observed in barrel }  \mid \text { $i$-th bin of $\cos\theta_{\bar{n}}$}) \\
		&= \frac{N_i^{\rm obs}}{N_i^{\rm gen}},
	\end{aligned}
\end{equation}
where $N_i^{\rm obs}$ is the number of events found in the barrel in the $i$-th bin of $\cos\theta_{\bar{n}}$ 
at the generator level, 
and $N_i^{\rm gen}$ is the number of events of the latter. 
For each event, we generate a random number $u_1$ from $\mathcal{U}(0, 1)$.
If $u_1 < R_{i}$, the anti-neutron should be considered as detected in barrel, otherwise it
should be considered as detected in end caps. 
Due to the limitation of statistics and the beam-associated background, the quality of the simualtion of the anti-neutron in end caps is worse than that in barrel, as shown in Fig.~\ref{fig:efficiency}.

In the sampling of kinematic variables, information of the anti-neutron, {\it e.g., $\Delta\alpha_{\rm gen}$}, from its generator level is needed. However,
this can only be obtained from the recoil system of $p\pi^{-}$ in data, $\Delta\alpha_{\rm recoil}$. 
The difference between $\Delta\alpha_{\rm recoil}$ and  $\Delta\alpha_{\rm gen}$ will lead to a bias in the simulation. 
To correct this,
$\Delta\alpha_{\rm recoil}$ can be considered as the convolution of $\Delta\alpha_{\rm gen}$  with a resolution function $g$,

\begin{equation} \label{eq:alpharecoil}
	\Delta\alpha_{\rm recoil} = \Delta\alpha_{\rm gen} \otimes g.
\end{equation}
An interface class $\sf TVirtualFFT$ provided by {\sc ROOT}~\cite{2017ROOT} is used to perform a 
deconvolution and to estimate the true four momenta of the anti-neutron.

\section{Comparison to experimental data}
\label{sec:comparison}

\subsection{$J/\psi\to p\bar{n}\pi^{-}$ }

The preliminary selection criteria of $J/\psi\to p\bar{n}\pi^{-}$ are the same as in Sec.~\ref{sec:datasample}.
Further requirements on the characteristic variables of the anti-neutron are applied, {\it e.g.} 
$E_{\bar{n}}>0.48$~GeV, $S_{\bar{n}}>18$~cm$^{2}$ and $H_{\bar{n}}>20$. The set of selection criteria is widely used in 
analyses to distinguish showers between neutron and anti-neutron.
The selection efficiency of the anti-neutron  with respect to  $\cos\theta_{\bar{n}}$ and  $p_{\bar{n}}$ as well the divergence between data and simulated MC are shown in
 Figures.~\ref{fig:efficiency}(a) and (b). Excellent agreement can be observed for 
data and the {\sc Geant4}-independent simulation.

Furthermore, a three-constraint kinematic fit imposing energy-momentum conservation is implemented under the hypothesis of $J/\psi\to p\bar{n}\pi^{-}$, and the corresponding $\chi^{2}$ is required to be less than 40.
Fig.~\ref{fig:efficiency}(c) and (d) show the efficiency after the applied kinematic fit,
where the overall difference between data and simulation is less than 1\% in $\cos\theta_{\bar{n}}$.
Due to the sensitivity of the kinematic fit and the low statistics  at low momentum of the anti-neutron  in $J/\psi\to p\bar{n}\pi^{-}$,
the difference is found to be larger than 1\% at low  $p_{\bar{n}}$,
 while the difference between data and simulation is less than 1\% for $p_{\bar{n}}>0.6$~GeV/$c$. 

\subsection{ $\Lambda\to \bar{n}\pi^{0}$ }

\begin{figure}[h]
 \centering
  \mbox
  {
  \begin{overpic}[width=0.45\textwidth]{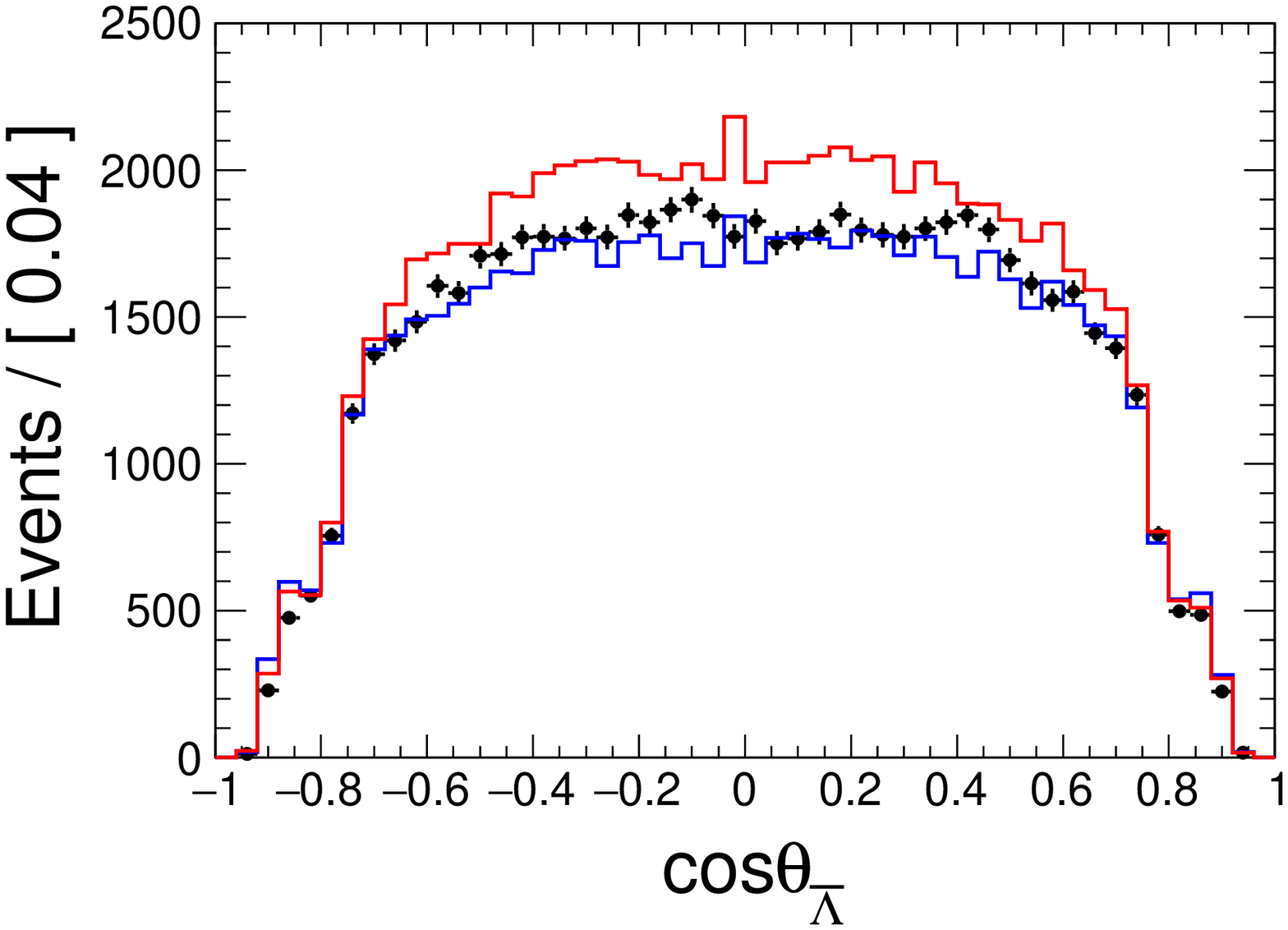}
  \put(80,55){$(a)$}
  \end{overpic}
  }
  \mbox
  {
  \begin{overpic}[width=0.45\textwidth]{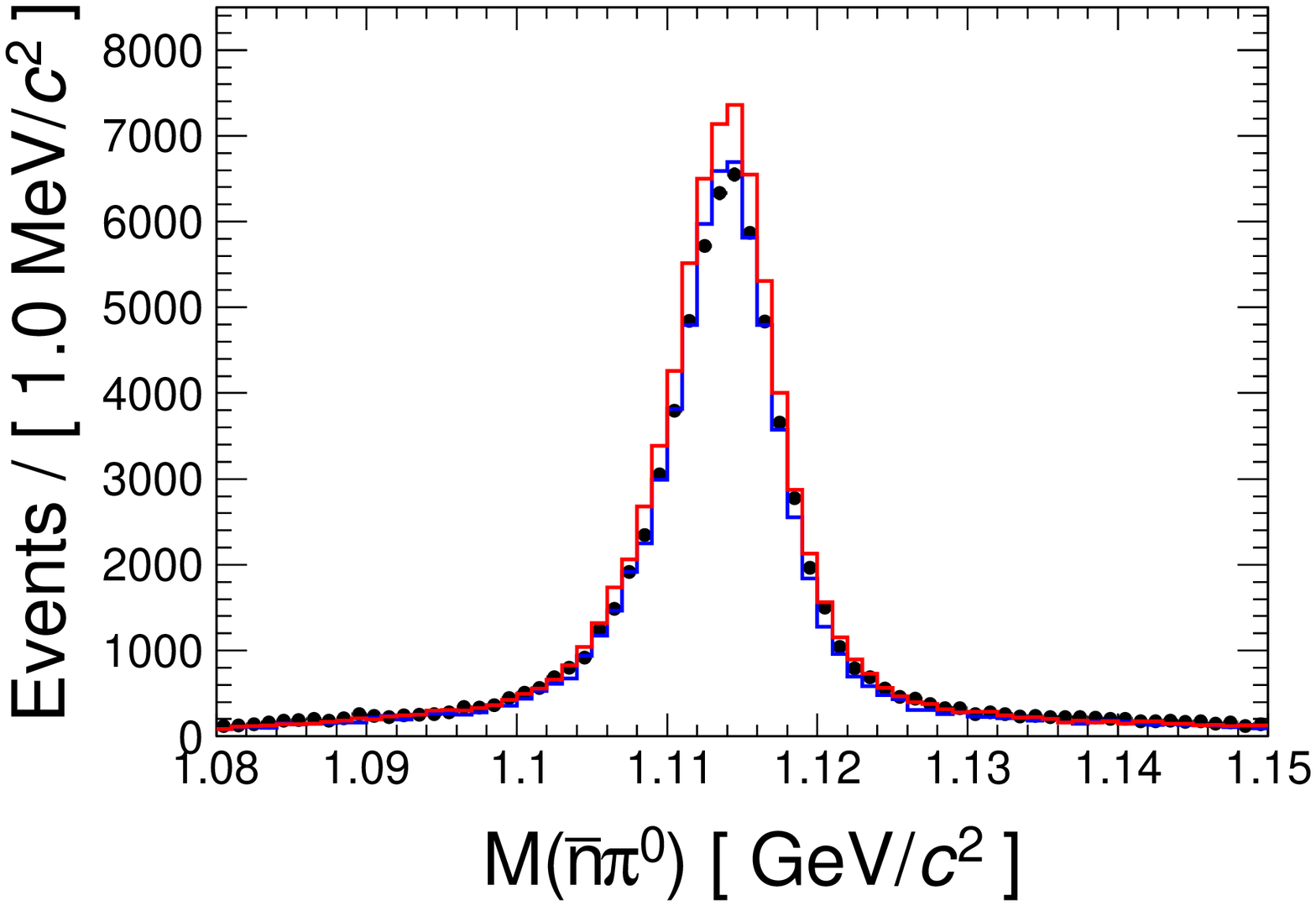}
  \put(80,55){$(b)$}
  \end{overpic}
  }
\caption{ (a) Angular distribution of $\bar{\Lambda}$ and (b) distribution of $\bar{n}\pi^{0}$ invariant mass.
	 Black dots with error bars show data;  red lines indicate weighted PHSP simulation which is {\sc Geant4}-beased;  blue lines represents the {\sc Geant4}-independent simulation with utilizing the data-driven method as described in this paper. The simulations are normalized according to the branching fraction of 
$J/\psi\to \Lambda\bar{\Lambda}$ according to Ref.~\cite{ParticleDataGroup:2020ssz}.
}
\label{fig:npi0}
\end{figure}

The decay process $J/\psi \to \Lambda \bar{\Lambda} \to p\pi^- \bar{n}\pi^0$ is used to study the 
polarization of $\Lambda$
and probe $CP$ violation~\cite{BESIII:2018cnd}. A precise simulation of anti-neutron related observables 
is essential for improving
the sensitivity of $CP$ violation at BESIII. 
Here we present the measurement of branching fraction of $\bar{\Lambda}\to \bar{n}\pi^{0}$
using the anti-neutron simulation developed in this work.
The process $\Lambda\to p\pi^{-}$ is denoted as {\it tag side} with following selection criteria:
candidates for $\Lambda$ are reconstructed by combining two oppositely charged tracks into the final states $p\pi^-$;
two daughter tracks are constrained to originate from a common decay vertex by requiring  $\chi^2$ of the second
vertex fit to be less than 35. The single tag efficiency $\varepsilon_{ST}$ is $54.82\%$ and the signal yield $N_{ST}$ is 
$(9.32\pm0.01)\times10^{5}$ utilizing 1.3 billion $J/\psi$ events collected at BESIII. 
For the {\it signal side} of  $\bar{\Lambda}\to \bar{n}\pi^{0}$,
at least three showers are required.
The most energetic shower with deposited energy larger than $\SI{0.4}{\GeV}$ is 
considered as the anti-neutron candidate. Another pair of showers, differing at least by $\ang{20}$ from the anti-neutron shower vector, are 
used to reconstruct the $\pi^0$ candidates. To improve the resolution, a three-constraint (3C) kinematic fit 
is performed imposing energy-momentum conservation, and the $\chi^2_{3C}$ is required to be less than 200. The double tag efficiency $\varepsilon_{DT}$ 
is $12.39\%$ and the final signal yield $N_{DT}$ is $(7.48\pm0.04)\times10^{4}$. The branching fraction 
$\mathcal{B}(\bar{\Lambda}\to \bar{n}\pi^{0})$ can be determined by 

\begin{equation}
\mathcal{B}(\bar{\Lambda}\to \bar{n}\pi^{0}) = \frac{N_{DT}\times\varepsilon_{ST}}{N_{ST}\times\varepsilon_{DT}\times\mathcal{B}(\pi^0 \to \gamma\gamma)},
\end{equation}
to be $(36.0\pm0.5)\%$ with the anti-neutron simulated using the data-driven method described in this paper.
The uncertainty is statistical.
The result is consistent with the PDG value $(35.8\pm0.5)\%$~\cite{ParticleDataGroup:2020ssz}
within the statistical uncertainty.
The distributions of angular position and the invariant mass of $\bar{n}\pi^{0}$ can be found in Fig.~\ref{fig:npi0}.

\section{Summary}
\label{sec:summary}

In this paper, a data-driven method developed to provide a {\sc Geant4}-independent simulation of  
the detector response of the anti-neutron in the EMC at BESIII 
is introduced. A large amount of anti-neutrons is 
selected from the process $J/\psi\to p\bar{n}\pi^{-}$. 
The performance of the anti-neutron in the EMC is simulated, including its detection efficiency
and various observables.
Our approach ensures a high agreement between data and simulation, when determining the detection 
efficiency of the anti-neutron, as well as a satisfying precision from a kinematic fit procedure, 
{\it e.g.} the systematic uncertainty can be reduced to be 1\% on average.
The algorithm can be used effectively in a variety of physics processes in the future at BESIII, such as 
 the measurement of nucleon electromagnetic form factors, or the hyperon decay and corresponding $CP$ sensitivity  probe of hyperons. 

We want to mention that this method can also be used to simulate other neutral particles at BESIII, such as the neutron and $K_{L}$.
At BESIII, which collected 10 billion $J/\psi$ events, a large amount of neutrons can be selected from $J/\psi\to \bar{p} n\pi^{+}$,
 and $K_{L}$ can be selected from  $J/\psi\to K_{L}K^{\pm}\pi^{\mp}$, respectively, which ensures the feasibility of simulating 
the detector response of these neutral particles with our data-driven method.

\section*{\boldmath Acknowledgments}
The authors gratefully thank the BESIII Collaboration for the profitable discussion and internal review,
thank Kai Zhu, Chunxiu Liu and Paul Larin for their suggestions and proofreading, thank Xiaobin Ji, Karin 
Schonning, Weimin Song, Bo Zheng, Liaoyuan Dong and Wolfgang Gradl for their help to improve the analysis. 
This work is supported partly by the National Science Foundation of China under Project No.12122509, 11911530140, U1832103, 11625523,
USTC Research Funds of the Double First-Class Initiative YD2030002005 and the Fundamental Research Funds for the Central Universities.

\bibliographystyle{elsarticle-num}
\bibliography{bibitem}

\end{document}